\documentclass[10pt,twocolumn]{article}
\usepackage{graphicx,epsf,bm,bbm,color,amsmath,ulem,amssymb,psfrag,float}
\usepackage{cuted}
\usepackage{hyperref}
\usepackage{balance}
\usepackage{authblk}

\usepackage{comment} 
	\newcommand{\mm}{m_e}
	
	\newcommand{\beq}{\begin{equation}}
	\newcommand{\be}{\begin{equation}}
	\newcommand{\beqn}{\begin{eqnarray}}
	\newcommand{\eeq}{\end{equation}}
	\newcommand{\ee}{\end{equation}}
	\newcommand{\eeqn}{\end{eqnarray}}

	\newcommand{\ep}{{\epsilon}}

		\newcommand{\rr}{{\omega}}

				 \newcommand{\osc}{\text{osc}}
\newcommand{\bem}{\begin{pmatrix}}
\newcommand{\eem}{\end{pmatrix}}

\newcommand{\red}{\color{red}}
\newcommand{\blue}{\color{blue}}
\newcommand{\ci}{\text{Ci}}
\newcommand{\si}{\text{Si}}

\newcommand{\limite}[1]{ {{\raisebox{-.3cm}{$\textstyle\longrightarrow$}} \atop {\scriptstyle{#1}}}}

\newcommand{\limiteup}[1]{
 {
 {\scriptstyle{#1}}
 \atop
{\raisebox{.3cm}{$\textstyle\longrightarrow$}}
}
}

\begin{document}

\title{Ramanujan, Landau and Casimir, \\
divergent series : a physicist point of view}

\author{Gilles Montambaux}

\affil{Laboratoire de Physique des Solides, Universit\'e Paris-Saclay, CNRS  UMR 8502 , 91405-Orsay, France}

\date{\today} 
\maketitle

\begin{abstract}

It is a popular paradoxical exercise to show that the infinite
sum of positive integer numbers is equal to $-1/12$, sometimes called the Ramanujan sum.
Here we propose a qualitative approach, much like that of a physicist, to show how the value  $-1/12$ can make sense and, in fact, appears in certain physical quantities where this type of summation is involved. 

At the light of two physical examples, taken respectively from condensed matter --the Landau diamagnetism-- and quantum electrodynamics --the Casimir effect-- that illustrate this strange sum,  we present a systematic way to extract this Ramanujan term from the infinity.

\end{abstract}

 \section{Introduction}

Consider the sum of natural numbers:
\be    s(n) = 1 + 2 + 3 + 4 + 5 +  \cdots + n  \ee
 For the first $n$ numbers, this sum has been calculated by young Gauss who found that
\be s(n) = {n (n+1) \over 2} \ee
which of course goes to infinity when  $n \rightarrow \infty$.
But certain ways to calculate this {\it infinite} sum lead to the   result written in a fancy form~:
\be    s(\infty) = 1 + 2 + 3 + 4 + 5 +  \cdots \longrightarrow   - {1 \over 12} \label{Rama1} \ee
How can the infinite sum of positive numbers be finite and negative?!

One can get this strange result by  a subtle manipulation of infinite sums, as proposed by Ramanujan (see a brief reminder in Appendix  \ref{app.Ramanujan}). The problem is that this manipulation is poorly defined: if one subtracts two infinite sums, what meaning does their difference have? This difficulty was highlighted by Ramanujan in his letter to Hardy\cite{Ramanujan}, and much earlier, this type of result had been obtained by Euler. 
In the following, we give the result of Eq.(\ref{Rama1}) and other infinite sums, the name of {\it Ramanujan sum} and will label them with the letter $\mathcal R$.
\medskip

 By appropriate regularisations, mathematicians have shown ways to give a correct signification to this strange result. A regularisation is precisely a way to define the infinity, and extract a finite number from this infinity. In some sense, this sum is equal to "infinity -1/12".  Furthermore, it  is related to the Riemann $\zeta$ function:
\be \zeta(x)= \sum_{n=1}^\infty {1 \over n^x} \ee
which is well-defined for $x>1$ but diverges otherwise. Riemann extended its definition to negative (and complex) arguments using a method called analytic continuation, yielding the result $\zeta(-1)=-1/12$.

\medskip

In this paper, we develop a tutorial presentation to understand the nature of this $-1/12$, by taking two examples of physical quantities where this kind of infinite sum appears.  One is the Landau diamagnetism of free electrons. The other is the energy of vacuum associated with the Casimir effect. 
 \medskip

Landau diamagnetism, the orbital reponse of a free electron gas to a magnetic field, is accompagnied by quantum oscillations when the field varies. The usual treatment of these oscillations provides an interesting clue to qualitatively understand the separation between the "`infinity"' and the $-1/12$ contribution. This is developed in the next section. In a magnetic field, due to the quantization of the electronic spectrum into discrete Landau levels, the variation of the total energy with the Fermi energy exhibits a step-like behavior. 
By analysis of this behavior and using the Poisson summation formula, one separates the energy into three parts: a zero-field energy, an oscillating term known as de Haas-van Alphen oscillations, and a non-oscillating term proportional to $1/12$. This last term, the Landau diamagnetism, is a manifestation of a {\it Ramanujan sum} revealed by the magnetic field and it shows how an infinite sum is involved in a physical phenomenon.
\medskip

Then we follow the same type of systematic analysis for various infinite series to extract their {\it Ramanujan sum}. Using the same recipe for various infinite series, we separate the infinity in two parts, a power law which tells us how the sum scales to infinity and a correction to this infinite sum identified with the Ramanujan sum. This is done in sections \ref{sect.integers}-\ref{sect.other}.
\medskip

 The Casimir pressure is an attractive force between two metallic plates resulting from vacuum fluctuations of the electromagnetic field.   
In section \ref{sect.Casimir}, we discuss the Casimir effect in which this kind of sum appears, first in a tutorial one-dimensional example, then in three dimensions.
\medskip

 In conclusion, we argue that the ''infinite'' sum appears to be a vacuum energy (the energy of the filled Fermi sea in the case of electrons), and the Ramanujan sum appears in the response to an external parameter, the magnetic field in the case of diamagnetism, the variation of thickness between the plates in the Casimir effect. We stress the pedagogical analogy between the two calculations.

\section{Landau diamagnetism}


\begin{figure}[h!]
\centerline{
 \includegraphics[width=8cm]{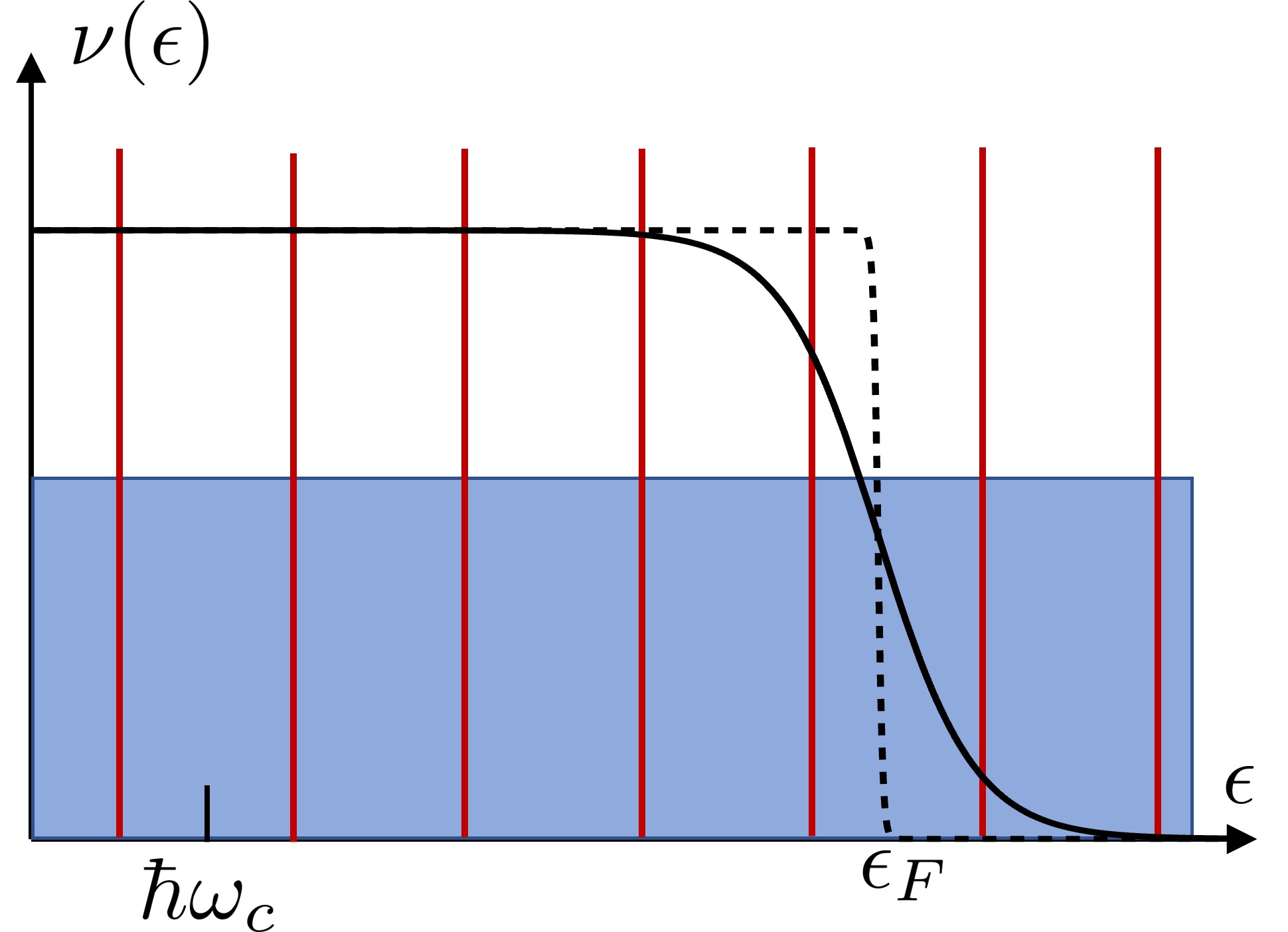} }
 \caption{ \small \it Density of states of $2D$ free electrons. Blue: zero field. In a magnetic field, the quantum states condense into Landau levels, of energy $(n+1/2)\hbar \omega_c$ and degeneracy $eB/h$.}
 \label{fig:spectre-landau}
\end{figure}
 
We consider the total energy of a two-dimensional ($2D$) electron gas. Due to Fermi principle, the electron states are populated up to an energy called the Fermi energy $\ep_F$ which depends on the electron density.
Figure (\ref{fig:spectre-landau}) recalls how the electronic states are populated. For each energy there is a finite and constant density of states given by (we do not consider spin degeneracy):
\be \nu(\ep) = {\mm \over 2 \pi \hbar^2}\ .  \ee
The total energy is therefore
\be E(\ep_F)= \int_0^{\ep_F} \nu(\ep)\,  \ep \, d\ep = {\mm \over 4 \pi \hbar^2 }\ep_F^2  \ .  \label{energy1}\ee

\begin{figure}[h!]
\centerline{
 \includegraphics[width=8cm]{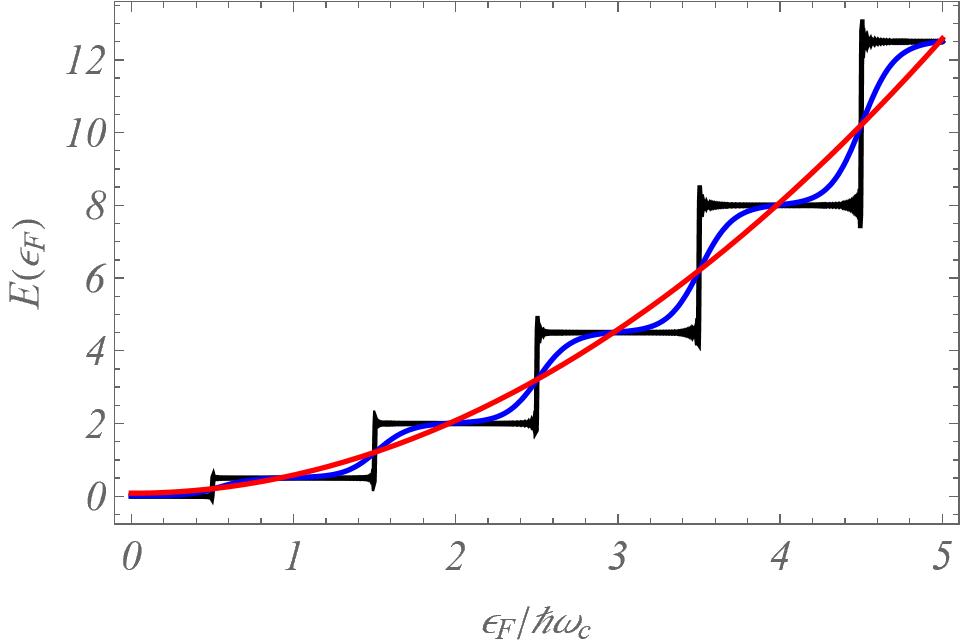 }}
 \caption{ \small \it Total energy of \, $2D$ electron gas in a magnetic field, as function of the ratio $\ep_F/\hbar \omega_c$. $T=0K$:  black steps. Finite temperature $k_B T \ll \hbar \omega_c$, blue curve. At temperature $T$ such that $\hbar \omega_c \ll k_BT \ll \ep_F$, one recovers the zero field quadratic increase, with a small field dependent correction, not visible at this scale (red curve).} 
\label{fig:energie-landau}
\end{figure}

In a magnetic field, Landau showed that the energy can take only a discrete set of values. Quantum states coalesce into the so-called Landau levels with energy (Fig. \ref{fig:spectre-landau}):
\be \ep_n = (n+{1 \over 2}) \hbar \omega_c    \ee
where $\omega_c = {e B / \mm}$ is the cyclotron frequency.
The total   density of quantum states being conserved, each Landau level has a degeneracy ${e B /h}$ per unit area.
The total energy now reads as a discrete sum~:

\be E(\ep_F)= {e B \over h} \left( 1 + 3+ 5 + 7 + 9 +\cdots \right) {\hbar \omega_c \over 2} \label{EeFB} \ee
with $\ep_n < \ep_F$.
Therefore we have to calculate the infinite sum of odd numbers (This is a problem slightly different from the one of the sum of natural numbers that will be considered below). 
The sum is actually not infinite but cut by the Fermi energy $\ep_F$. We are interested in the dependence  $E(\ep_F)$  that we write in the form:

 \be E(\ep_F)= {\mm \omega_c^2 \over 2 \pi} \sum_{n \geq 0} (n+{1 \over 2}) \, \Theta[ \ep_F - (n+{1 \over 2}) \hbar \omega_c] \label{energyT0} \ee
where $\Theta(x)$ is the Heaviside function. As seen on Figure (\ref{fig:energie-landau}), this sum has a step-like dependence in the Fermi energy (experimentally, the Fermi energy is rather fixed and the oscillations are seen when varying the magnetic field).  This step-like behavior leads to oscillations of the magnetization known as de Haas-van Alphen oscillations \cite{Landau,Shoenberg}.


To highlight and characterize these oscillations, it is common to use the Poisson summation formula (Appendix \ref{app.Poisson}) to obtain:\footnote{Textbooks usually rather consider the grand potential at finite temperature. The energy can be deduced with appropriate derivative of the grand potential \cite{Shoenberg}.}
\begin{eqnarray}
& &E(\ep_F) = {\mm \omega_c^2 \over 2 \pi  }  \left({1 \over 2} \left({\ep_F \over \hbar \omega_c}\right)^2 \right.  \\
&& +  \left. \sum_{m \geq 1} (-1)^m {\cos 2 \pi m x + 2 \pi m x \sin 2 \pi m x {\red -1} \over 2 (\pi m)^2} \right) \nonumber
\label{EeFpoisson}
\end{eqnarray}
where $x=\ep_F/\hbar \omega_c$. Examination of this formula shows three types of contribution~: a) The first term is the zero field total energy. b) The oscillatory terms are signatures of the step-like form of  the total energy vs. Fermi energy.  c) There is an additional non-oscillating  term,  dependent on the magnetic field, proportional to the sum

\be {\red {1 \over 2} \sum_{m = 1}^\infty {(-1)^{m+1}  \over  (\pi m)^2 } = {1 \over 24}  \ . }\ee
We can summarize the result in the form:

\be E(\ep_F)= {\mm \ep_F^2  \over 4 \pi \hbar^2 } + {e^2 B^2 \over 2 \pi \mm
} \left({1 \over 24}+ \osc(\ep_F / \omega_c) \right)\ee
where $\osc(x)$ scales as $x$ and $\langle \osc(x)/x \rangle=0$. The average is taken over a large ($\gg 1$)  range of $x$.
 As we recall later, the quantum oscillations are damped by temperature or disorder, so that only the two first terms remain.  The Landau susceptibility is given by the second derivative of the energy with respect to the magnetic field:

\be \chi = -{\partial^2 E \over \partial B^2} = -{e^2 \over 24 \pi \mm } \ .  \label{Landauchi} \ee

We see clearly that this diamagnetic term has is origin in an infinite series (\ref{EeFB}) of odd numbers,  very similar to the series (\ref{Rama1}), the main subject of this paper.
One can say here that the Ramanujan sum of this infinite series is revealed by the magnetic field. Indeed there are two characteristic energy scales in this problem, the Fermi energy which drives the infinity and the cyclotron energy which gives the correction to this infinity.
This physical separation will be very useful in the following discussion.
Taking  $\hbar=1$,  $\mm=2 \pi$, and $\omega_c=b$, the sum (\ref{EeFB}) gets the form, omitting the oscillations:
 \be   b^2 ({1 \over 2} +{3 \over 2}+{5\over 2}  +\cdots+{\ep \over b}) \  \limite{b \rightarrow 0}\  {\ep^2 \over 2}+ {b^2 \over 24} \ .  \label{epb} \ee

The notation (\ref{epb}) is precious for physicists because we see that, with the help of {\it two characteristic scales}, we  describe both the behavior at infinity and the Ramanujan sum. As we see now,   a third scale is needed to properly suppress the oscillations.
 \medskip

\noindent
{\it -- Damping of the oscillations --}
\medskip

The oscillations occur because of the sharp cut-off at the Fermi energy, when the temperature $T=0$. At finite temperature, this discontinuity is smoothed and the Heaviside function is replaced by a Fermi-Dirac factor:\footnote{Actually $\chi_\beta(\ep_F -\ep_n)=1 - f(\ep_n)$ where $f(\ep_n)$ is the Fermi factor with an inverse temperature $\beta$ and a Fermi energy $\ep_F$.}

\be \Theta(\ep_F -\ep_n) \longrightarrow   
\chi_\beta(\ep_F -\ep_n) \equiv {1 \over e^{\beta(\ep_F-\ep_n)}+1} \ee
where $\beta=1/k_B T$ with $k_B T \ll \ep_F$.
As temperature increases, the oscillatory terms in the sum (\ref{EeFpoisson}) are progressively reduced and disappear when $\omega_c \ll 1/\beta$ (Fig. \ref{fig:energie-landau}). For example, at finite temperature, 

\be  \cos 2 \pi m {\ep_F \over \omega_c} \longrightarrow  R(Z_m) \  \cos 2 \pi m {\ep_F \over \omega_c} \ee
where the reduction factor $R(Z_m)$ is given by
\be R(Z_m)= {\pi Z_m \over \sinh \pi Z_m}    \qquad \text{with} \  Z_m={2 m \pi^2 \over \beta \omega_c} \ . \ee
 Note that the power-law and the susceptibility are not affected by the temperature as long as $1/\beta \ll \ep_F$.
As discussed in ref. \cite{Shoenberg} and in Appendix \ref{app.damping}, 
the oscillations can be damped by other factors like   impurity scattering .

\medskip

\noindent
{\it -- Conclusion of this section --}
\medskip

We summarize the procedure used to extract the Landau susceptibility as a Ramanujan sum. We have replaced the sum over Landau levels (odd numbers) by a step function. This step function has three components, a power law which tells how the sum scales when the Fermi energy goes to infinity, an oscillatory behavior, and the field dependent (but $\ep_F$ independent) term which corresponds to a Ramanujan sum. In dimensionless units, we transform Eq. (\ref{epb}) and we conclude that the sum of odd numbers scales as:

\be 1+3+5+7+\cdots+ x \ \limite{x \rightarrow \infty} \\ {\blue {x^2  \over 4}} + {\red {1 \over 12}}  \ . \label{odd-numbers} \ee

\noindent
In the following, we apply the same procedure to study various infinite sums.

\medskip

\section{Infinite sum of integer numbers}
\label{sect.integers}


\begin{figure}[h!]
\centerline{
 \includegraphics[width=8cm]{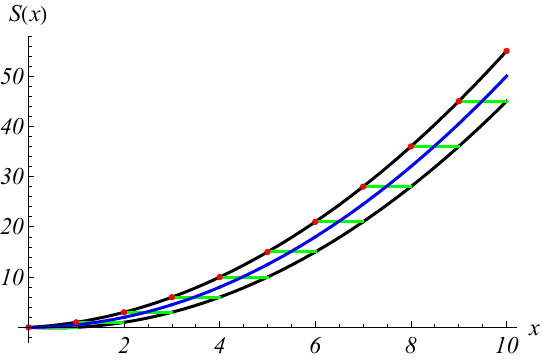} }
 \caption{ \small \it  Red dots: discrete values $s(n)$. Green: the step function $S(x)$. Black: the functions $x(x+1)/2$ and $x(x-1)/2$. Blue: the function $x^2/2-1/12$.} \label{fig:sn}
\end{figure}

We now come to our original problem. Consider the sum of natural numbers. The same procedure as above is used to show that, in the sense of Ramanujan:
\be    s(\infty) = 1 + 2 + 3 + 4 + 5 +  \cdots \limiteup{\mathcal R}\   - {1 \over 12}  \ . \ee

Inspired by the Landau problem, we define a step  function $S(x)$ that jumps to the value $s(n)$ for each integer $n$. This function $S(x)$ is an oscillating function that oscillates between the values $x(x+1)/2$ and $x(x-1)/2$ (see Figure \ref{fig:sn}). How can we define the average behavior of $S(x)$ as $x$ tends to infinity?
For example, this could be the average between the two continuous curves, i.e., $x^2/2$. A systematic approach proposed here is to decompose this oscillating function into harmonics as in the Landau calculation and then remove all harmonics to obtain an average value $\langle S(x) \rangle$. 
The function $S(x)$ is expressed as:

\be S(x)= \sum_1^\infty n \ \Theta(x-n)  \label{n-theta}\ee 
where $\Theta(x)$ is the Heaviside step function.

\begin{figure}[h!]
 \includegraphics[width=6cm]{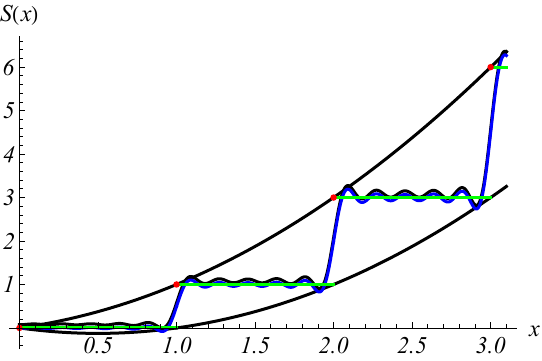} 
 \includegraphics[width=6cm]{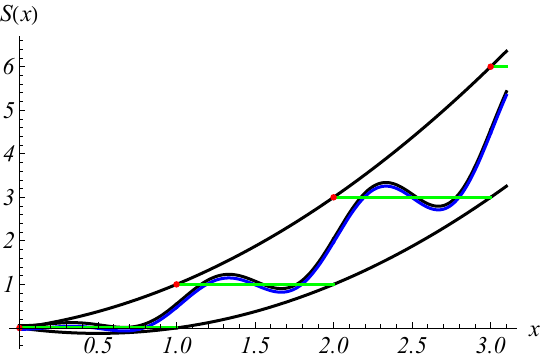} \\
	  \includegraphics[width=6cm]{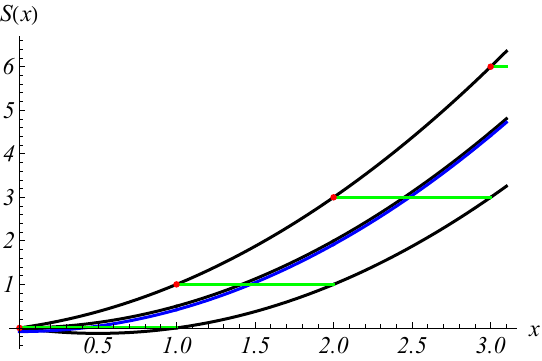} 
 \caption{ \small \it Fit of the function $S(x)$ by the Fourier series, limited here to $5$ terms, then to a single term, and then with no harmonics. In black, without the factor $-1/12$; in blue, with the factor $-1/12$ (the difference is barely visible).
} \label{fig:sn2}
\end{figure}

The Fourier transform allows us to express it as a sum of periodic functions. This way, we can separate an average behavior that tells us how this function behaves as it tends to infinity, from an oscillating behavior. Poisson's summation formula enables us to show that (see Appendix \ref{app.Poisson})

\be S(x)= {\blue{x^2 \over 2}} + 2 \sum_{m=1}^\infty \left[{x \sin 2 \pi m x \over 2 \pi m}+{\cos 2 \pi m x {\red - 1} \over (2 \pi m)^2 }   \right] \label{F1} \ee
The first term tells us how the function behaves as it tends to infinity. In addition, we see that in this sum, there exists a non-oscillating contribution:
\be {\red - 2 \sum_{m=1}^\infty  {1 \over (2 \pi m)^2} =  -{1  \over 12}}  \ee
which is the Ramanujan sum of naturals. 
Thus, $S(x)$ is the sum of a monotonic average contribution and an oscillating function $\osc(x)$:

\be \boxed{S(x)= {\blue{x^2 \over 2}}{\red  - {1 \over 12}} + \osc(x)}\ee
which is shown on Fig.\ref{fig:sn2}. We have learned in the previous section that the oscillations can be removed by replacing the Heaviside function $\Theta(x-n)$ by a smoothed fonction $\chi_\beta(x-n)$ with $1 \ll 1/\beta \ll x$. 
\medskip

For completion, we recall here  the sum of odd numbers, obtained from the Landau calculation (\ref{odd-numbers}).
\medskip

\centerline{
$\boxed{{1 } +  {3 } + {5 }  + {7 } + \cdots \limiteup{\mathcal R} \  \red {1 \over 12}}$}
\medskip

\noindent
and the sum of even numbers is obviously
\medskip

\centerline{
$\boxed{{2 } +  {4 } + {6 }  + {8 } + \cdots \limiteup{\mathcal R} \ \displaystyle{\red -{1 \over 6}}}$}
\medskip



\section{Sum of squares, even powers}
\label{sect.squares}

We   now proceed similarly for the sum of squares to  show that:

\be s_2= 1 + 4 + 9 + 16 + 25 + \cdots \limiteup{\mathcal R} \ {\red 0} \  \ !! \ee
Let's see how our method works.
Following the same procedure as before, we calculate the function 
\be S_2(x)= \sum_{n=1}^\infty n^2 \ \Theta(x-n) \ . \label{S2x} \ee
Its  Fourier decomposition is
\begin{eqnarray}
 &&S_2(x)={\blue {x^3 \over 3}}\label{F2} \\
 && +2 \sum_{m=1}^\infty \left[{\sin 2 \pi m x \over 2 \pi m } + {2 x \cos 2 \pi m x \over (2 \pi m)^2}-2 {\sin 2 \pi m x \over (2 \pi m x)^3} \right] \nonumber \end{eqnarray}
We see that this sum does not contain a constant term~:
\be \boxed{S_2(x) = {\blue{x^3 \over 3}} + {\red{0}} + \osc(x)} \ee
In the same sense as above, one can say that the Ramanujan sum  $s_2(\infty)={\red 0}$ !!
\medskip

Using the same procedure, one can show that all sums of even power of integer numbers are zero in the sense of Ramanujan. As recalled in Appendix \ref{app.Poisson}, this corresponds to the trivial zeros on the zeta function.

\section{Sum of odd powers}
\label{sect.odd}

Let's now show that:
\be s_3= 1 + 8 + 27 + 64 + 125 + \cdots \limiteup{\mathcal R}  {\red {1 \over 120}} \ !! \ee
As we will see in section \ref{sect.Casimir}, it is this sum that is involved in the Casimir effect. We calculate the function
 \be S_3(x)= \sum_{n=1}^\infty n^3 \  \Theta(x-n) \ . \label{S3x} \ee
Its Fourier expansion, which contains oscillating terms that we do not detail (see 
Appendix \ref{app.Poisson}) and two non-oscillating terms, is of the form
 
\be S_3(x) = {\blue {x^4 \over 4}} + {\red   \sum_{m=1}{12 \over (2 \pi m)^4}} + \osc(x)=  {\blue {x^4 \over 4}} {\red + {1 \over 120}} + \osc(x)  \label{F3}\ee
Similarly

\be S_5(x) = {\blue {x^6 \over 6}} {\red -  \sum_{m=1}{240 \over (2 \pi m)^6}} + \osc(x)=  {\blue {x^6 \over 6}} {\red - {1 \over  252 }}+\osc(x) \label{F5}\ee
and

\be S_7(x) = {\blue {x^8 \over 8}} {\red +  \sum_{m=1}{10080 \over (2 \pi m)^8}} + \osc(x)=  {\blue {x^8 \over 8}}{\red  + {1 \over 240}} + \osc(x) \label{F7}\ee

\noindent 
which generalizes, for an odd power $p=2 k -1$~:

\be S_{2 k -1}(x) = {\blue {x^{2 k}  \over 2 k }} +{\red (-1)^k \sum_{m=1}^\infty {2 (2k-1)!\over (2 \pi m)^{2k}}} + \osc(x) \ee
The constant term is nothing but the $\zeta$ function taken in $1-2k$.

\be \boxed{S_{2 k -1}(x) = {\blue {x^{2 k}  \over 2 k }} + {\red \zeta(1-2 k)} + \osc  }\ee
\medskip

\noindent
We can now summarize our result obtained for even or odd powers of naturals in the form:
\be \boxed{S_{p}(x) = {\blue {x^{p+1}  \over p+1 }} + {\red \zeta(-p)} + \osc(x)  }
\label{Spx} \ee

\section{Other sums}
\label{sect.other}

We now proceed similarly to calculate sums which are not convergent and ill-defined. Then we reconsider the case of two convergent series.

\subsection{Infinite sum of ones}
\centerline{
$\boxed{{\boldsymbol s_0= 1 + 1 + 1 + 1 + 1 + \cdots \limiteup{\mathcal R} \ {\red -{1 \over 2}} \ . }}$}
\bigskip

Again, how the sum of positive numbers can be negative!
\medskip

We define the function $S_0(x)$, and we seek its Fourier expansion to extract its average behavior:
\be S_0(x)= \sum_{n=1}^\infty \Theta(x-n)= {\blue x} {\red -{1 \over 2}} +  \sum_{m=0}^\infty  { \sin 2 \pi m x \over  \pi m} \ . \label{S0x}  \ee
The Ramanujan  sum of ones is $-1/2$ 
 \medskip

\subsection{Alternating sum of integer numbers}

\centerline{
$\boxed{1-2+3-4+5-6+\cdots \limiteup{\mathcal R} \ \displaystyle{\red {1 \over 4}}}$
}
\bigskip

Here, we consider the alternating sum of the integers, denoted as $B$ in Appendix \ref{app.Ramanujan}:

\be B= 1 - 2 + 3 - 4 + 5 - 6 + \cdots  \ee
We introduce the function:
\be B(x) = -\sum_{n=1}^\infty  n \, \cos(n\pi)\,  \Theta(x-n)  \ . \label{Bdexdef} \ee
By retaining only the non-oscillating terms of the Fourier expansion, we find:

\begin{eqnarray} B(x)\! &=& \! {\red {1 \over \pi^2}   \sum_{{m=1 \atop k=\pm 1}}^\infty {1 \over (2m+k)^2}}+ \osc(x) \nonumber\\
&=& \!  {\red{1 \over 4}}+ \osc(x) \ . \end{eqnarray}

\noindent
The alternate sum of integers is $1/4$.
 \medskip

\subsection{Grandi series}

The so-called Grandi's sum is:
\be A= 1 - 1 + 1 - 1 + 1 - 1 + \cdots  \ee
We introduce the function: 
 \be A(x) = \sum_{n=1}^\infty \cos n\pi    \, \Theta(x-n) \ee whose Fourier expansion is:
 
\begin{eqnarray} A(x)&=& {\red  {1 \over 2}} +  {\sin \pi x \over \pi} + \sum_{m=1 \atop k=\pm 1}^\infty {\sin (2 m +k) \pi x \over (2m+k) \pi}\nonumber\\
&=& \! \!  {\red {1 \over 2}}+ \osc(x) \ . \end{eqnarray}
so that we can write:
\medskip

\centerline{$\boxed{1 - 1 + 1 - 1 + 1 - 1 + \cdots \limiteup{\mathcal R}\  \displaystyle{\red {1 \over 2}}}$}

\subsection{Harmonic series}
We continue to use the same method to calculate the sum of the  series:
\medskip

\centerline{
$\boxed{1 + {1 \over 2} + {1 \over 3}+ {1 \over 4} + \cdots \limiteup{\mathcal R}\ \displaystyle{{\blue \ln x}+{\red \gamma}}}$
}
\medskip

\noindent
where  $\gamma$ is the Euler constant.
We introduce the function:
\be S_{-1}(x) = \sum_{n=1}^\infty {1 \over n}    \, \Theta(x-n)\ee whose Fourier expansion is:

\be S_{-1}(x)= {\blue \ln x} + {1 \over 2} + 2 \sum_{m=1}^\infty [\ci(2 m \pi x) - \ci(2 m \pi)] 
\ee
 where $\ci$ is the cosine integral function. It satisfies the sum  $\sum_{m=1}^\infty \ci(2 m \pi)= 1/4 - \gamma/2$ . Therefore,

\be S_{-1}(x)= {\blue \ln x
}+  {\red \gamma}+ \osc(x) \ee

\subsection{The Basel problem}

\centerline{$\boxed{1 + {1 \over 2^2} + {1 \over 3^2}+ {1 \over 4^2} + \cdots = \displaystyle{\red {\pi^2 \over 6}}}$}
\bigskip

\noindent
We finish with the famous  "Basel problem" studied by Euler. Here, the infinite sum is convergent. Can we still use our method?

\medskip

 \noindent 
We introduce the function:

\be S_{-2}(x) = \sum_{n=1}^\infty {1 \over n^2}    \, \Theta(x-n)\ee whose Fourier expansion is:
 
\begin{eqnarray}
 &&S_{-2}(x)={3 \over 2}-{1 \over x}  \\
 &&\hspace{-0.5cm} + 2 \sum_{m=1}^\infty \left(1\! - \! {\cos(2 \pi m x) \over x} + 2m \pi [\si(2 m \pi)- \si(2 m \pi x)]\right) \nonumber
\end{eqnarray}
\noindent
where $\si$ is the sine integral function which verifies  $\si(2 m \pi x) \longrightarrow {\pi \over 2}$ si $x \rightarrow \infty$ . Therefore, $S_{-2}$ converges to the value:

\be S_{-2}(\infty)= {3 \over 2}+ 2 \sum_{m=1}^\infty [1  + 2m \pi \si(2 m \pi)-m \pi^2 ] \ee
\noindent
and we  have checked (numerically) that this sum is indeed ${\pi^2 / 6}$.

 \section{Back to physics~: Casimir force}
\label{sect.Casimir}
We now come back to physics, with a brief description of the Casimir effect and its connection with a Ramanujan sum.
\subsection{Introduction}
The Casimir effect is a manifestation of vacuum fluctuations of the electromagnetic field. 
The total energy of the field is a sum over electromagnetic modes of frequencies $\omega$:
\be E= \sum_{\text{modes}} \left( n(\omega,T) +{1 \over 2}\right) \hbar \omega \ . \ee
The first term
 is the contribution of the modes excited at finite temperature. $n(\omega,T)$ is the Bose factor. The summation over modes leads to the well-known result, per volume unit:
\be E_1(T)= {\pi^2 \over 15} {(k_B T)^4 \over (\hbar c)^3} \ .  \ee
This is the well-known energy of the black-body, which vanishes at zero temperature. The second term is more delicate since it is an infinite contribution of modes. It is thus infinite (and temperature independent). We now consider this second term, first in a rather academic $1D$ geometry, then in three dimensions.

\begin{figure}[h!]
 \centerline{\includegraphics[width=4cm]{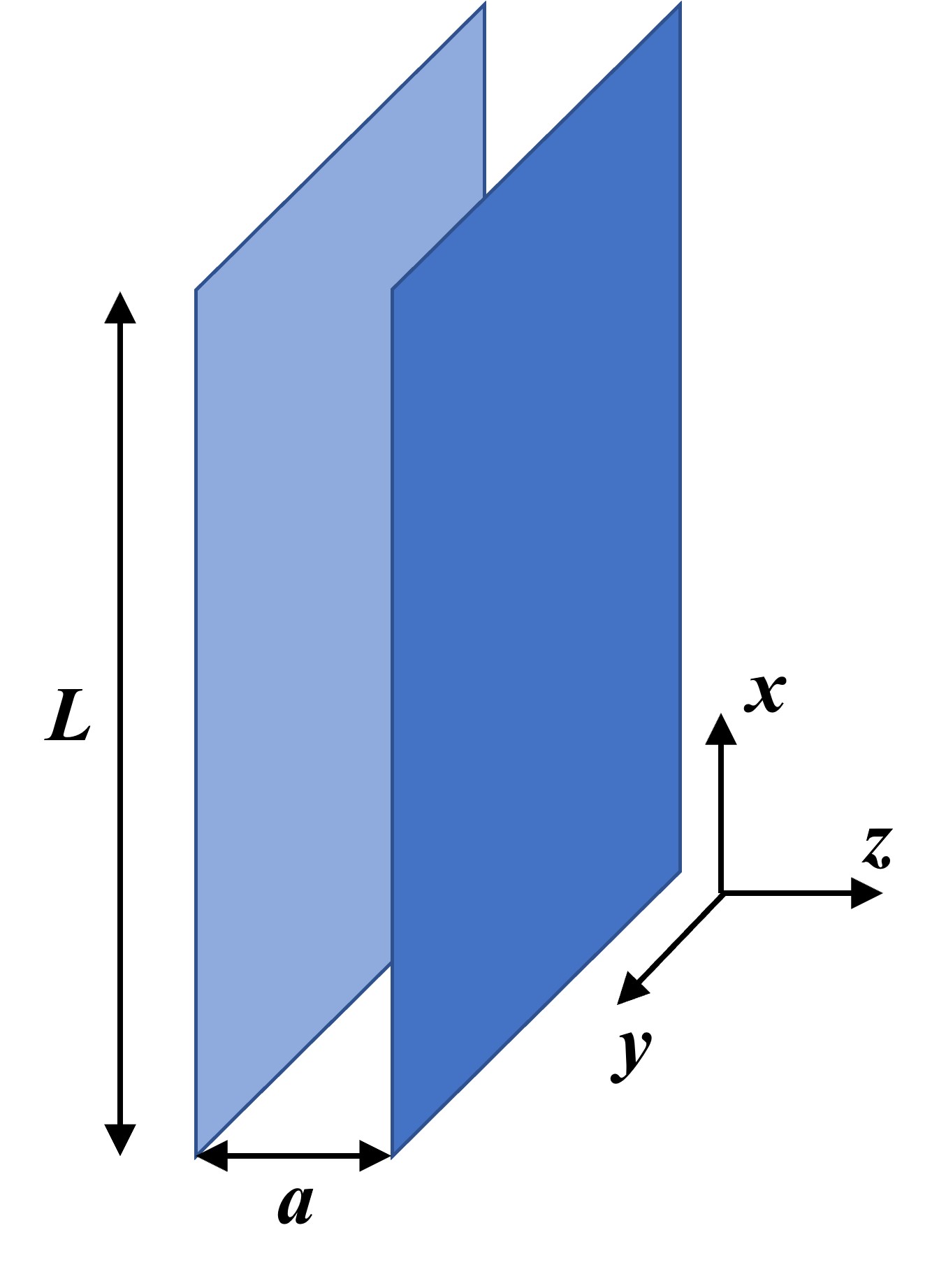}} 
 \caption{ \small \it Geometry of the Casimir effect. Two metallic plates at short distances are attracted by a small force.
} \label{fig:Casimir1}
\end{figure}

\subsection{Casimir effect in one dimension}

It is instructive to start with a 1D calculation, since it is 
very similar to the Landau calculation. We consider the space confined between two points distant of $a$. The sum is given by 

\be 
E={1 \over 2} \sum_{n=0}^\infty \hbar \omega_n \ee
where the frequency of the modes confined between the two points   is $\omega_n= n \pi c /a$. This sum is in principle infinite, but Casimir considers that if the confining points are metallic, modes with a frequency larger that a characteristic  frequency cannot be confined \cite{Casimir}. This is the plasma frequency (this will be more clear and physical in $3D$). The sum is actually finite, limited to a characteristic frequency $\omega_p$.

 Therefore this maximal frequency $\omega_p$ drives the ''infinite'' sum, exactly as the Fermi energy controls the infinite sum of Landau levels in (\ref{energy1}). And the frequency of the lowest mode, $\pi c / a$ plays the role of the cyclotron frequency. We have the analogy:

\begin{eqnarray}
 \ep_F  &\longleftrightarrow& \hbar \omega_p  \nonumber \\
  \omega_c &\longleftrightarrow& {\pi c \over a} 
 \end{eqnarray}
We first assume that there is a sharp cut-off at frequency $\omega_p$, so that we write the sum as
\be 
E(\omega_p)={\hbar \pi c \over 2 a} \sum_{n=0}^\infty  n \,  \Theta(\omega_p - n \pi c /a)   \ . \ee
This sum on integers  has been calculated in (\ref{n-theta}). With the physical parameters of this problem, the Poisson summation gives:

\be 
E(\omega_p)={\hbar  a \over 4 \pi c}    \omega_p^2  - {\pi \hbar c \over 24 a}+ \osc\left({\omega_p a\over \pi c}\right)    \ee
In this form, the analogy with the Landau calculation is clear. Here the sum is over integers, giving $-1/12$, while for Landau it was over odd numbers, giving $1/12$.
As we discuss in more details in the next subsection, the first term is the vacuum energy, the second term leads to the Casimir force

\be F=-{\partial E \over  \partial a}= -{\pi \hbar c \over 24 a^2}\ee
which is the equivalent of Eq. (\ref{Landauchi}) for the Landau susceptibility.\footnote{The Landau diamagnetic energy is a quadratic function of $\omega_c$ while the Casimir force is a linear function of $\pi c/a$. The difference comes from the Landau degeneracy which gives a multiplicative power of $\omega_c$.}

\subsection{Casimir force in 3D}

After this exercise, we come now to the original paper of Casimir\cite{Casimir,Duplantier} who considers two large plates of size $L$ at short distance $a$ ($L \gg a$), Fig. (\ref{fig:Casimir1}). The total energy at $T=0$ is now 

\be 
E={\hbar  \over 2} \sum_{n,k_x,k_y} \omega_n(k_x,k_y)     \ee
where the frequency of the $3D$ modes is now  $\omega_n(k_x,k_y)= c  \sqrt{{n^2 \pi^2 \over a^2} +k_x^2+k_y^2}$ .  Following Casimir, we notice that this sum must have an upper cut-off since modes with frequency larger than the plasma frequency $\omega_p$ are not confined between the plates. Since $L$ is large, the transverse modes form a continuum and their sum can be replaced by an integral:

\be 
E(\omega_p)={L^2 \hbar c  \over 2 \pi}  \sum_{(n)}  \int_0^{k_p} \sqrt{{n^2 \pi^2 \over a^2} +k^2} \, k dk \label{integral-transverse}\ee
where the upper limit $k_p(n)$ on the  transverse vector $k$ is such that ${n^2 \pi^2 \over a^2}+ k_p^2 < {\omega_p^2 \over c^2}$. We have inserted a factor $2$ to account for the modes degeneracy. The discrete sum $(n)$ means that the lowest mode has to be counted only once, so it will be affected by a factor $1/2$. Integration over transverse momenta gives

 \be E(\omega_p)={L^2 \hbar c \over 6 \pi} \sum_{(n)} \left( {\omega_p^3 \over c^3} - {n^3 \pi^3 \over a^3}\right) \Theta\left({\omega_p a \over \pi c} - n \right) \label{sum-casimir} \ee
which we rewrite in a dimensionless form:

 \be E(\omega_p)={L^2 \pi^2 \hbar c \over 6   a^3}\left( \sum_{n=1} ( x^3 - n^3) \Theta(x-n )+x^3/2 \right) \ee
with $x= {\omega_p a \over \pi c} $. The last term is the lowest mode contribution. The sum is given in Appendix \ref{app.Poisson}. It is written in term of the sums discussed in the previous sections. It involves the infinite sum over cubes of integers calculated in (\ref{F3}):\footnote{\label{fnminus}Notice that the sum over cubes is preceded by a (-1) factor.} 

 \be E(\omega_p)={L^2 \pi^3 \hbar c \over 6 \pi a^3}\left( x^3 S_0(x) - S_3(x)+x^3/2 \right) \ee
where $S_0(x)$ and $S_3(x)$ are given by (\ref{S0x},\ref{F3}).
The term in parenthesis is 

\be {3 \over 4} x^4 -{1 \over 120}+ \osc(x) \ee
and the total energy is written as

 \be E(\omega_p)=L^2 a {\hbar \omega_p^4 \over 8 \pi^2 c^3} - L^2{ \pi^2 \hbar c \over 720 a^3} + \osc\left({\omega_p a \over \pi c}\right) \ . \ee
The first term is the vacuum energy. It is extensive (i.e. proportional to $L^2 a$ and the energy density is
\be 
E_0(\omega_p)= {\hbar \, \omega_p^4 \over 8 \pi^2 c^3} \ .\ee
The nature of this vacuum energy is subtle and will not be discussed here. However since this energy density exists also outside the plates, it does not contribute to any force on the plates.\footnote{This term is often not considered and not discussed in the literature, including the Casimir paper.} The second term scales with the aera of the plates. It is at the origin of the (negative) Casimir pressure which represents the attractive force between the two plates:\footnote{It is interesting to point out that the calculation of the Casimir force involves the zeta function $\zeta(-1)$  in $1D$ and $\zeta(-3)$ in $3D$. They have opposite signs, while the force is attractive in both cases. This is because the integral (\ref{integral-transverse}) over transverse modes induces a change in sign on the sum over cubes (\ref{sum-casimir}), see footnote \ref{fnminus}.}
\be P= -{1 \over L^2} {\partial E \over \partial a}= - {\pi^2 \over 720} {\hbar c \over a^4} \ee

Just as the Landau diamagnetism is the physical quantity which reveals the infinite sum over integer numbers, the Casimir pressure is the physical quantity which reveals the sum over cubes.

\section{Conclusion}
\label{sect.conclusion}

We have proposed an intuitive method to give a simple meaning to apparently strange terms which can be extracted from infinite non convergent sums, like the $-1/12$ for the infinite sum of naturals.

Based on the calculation of quantum oscillations and the diamagnetic response in a free electron gas in a magnetic field, we propose the simple scheme:
Replace the sum by a step function which takes its value at each number of the sum, Fourier transform this function to get 1) a principal term which tells how the sum scales to infinity, 2) oscillations which cancel in average and can be suppressed by  appropriate smoothing the step function, 3) a constant terms which remains when the oscillations have been suppressed. The latter term is precisely the {\it Ramanujan sum}. 

The two physical examples in which the Ramanujan sum is revealed by a physical quantity are interesting because both the scaling at infinity and the Ramanujan term have a physical meaning.
The first term is ''a vacuum'' energy. 
This is obvious in the Casimir example, because it results from the summation of the zero point energies of the harmonic oscillators which are the modes of the electromagnetic field. Its signification is complex and will not be discussed here, but dimensional analysis shows that it scales like the $4^{th}$ power of some frequency (or energy) scale. 

In the Landau problem, the first term  can also be considered as a ''vacuum energy'' since it is the total energy of a filled Fermi sea, which can be considered as the vacuum of electron-hole excitations. In three dimensions, it scales like the $2^{nd}$ power of the Fermi energy.
We can summarize these two physical points in the synthetic form:

\begin{eqnarray}
 1+3+5+7+\cdots+ x \longrightarrow {{\blue x^2 \over 2}}&+&{{\red 1\over 12}}+ \osc(x) \nonumber \\
\text{vacuum} &+& \text{Landau} \nonumber \\
1+2^3+3^3+\cdots+ x \longrightarrow {{\blue x^4 \over 4}}&+&{{\red 1\over 120}}+ \osc(x) \nonumber \\
\text{vacuum} &+& \text{Casimir} \nonumber
\end{eqnarray}

 \clearpage

\appendix

\section{Strange manipulations of infinite sums} 
\label{app.Ramanujan}

We recall here the Ramanujan calculations, taken from Wikipedia\cite{Ramanujan}.
\medskip

Let $A, B$ and $S$ three distinct sums, with  $S$ being the sum of natural numbers, such that:

$$   {\displaystyle A=1-1+1-1+1-\cdots} \qquad  \text{(Grandi series )}$$
   $$    {\displaystyle B=1-2+3-4+5-\cdots}  \qquad  \text{ (alternate sum )}$$
  $$     {\displaystyle S=1+2+3+4+5+\cdots} \qquad  \text{ (natural numbers)} $$
\medskip

$\bullet$ {\bf Determination of $A$}
\medskip 

One defines:
 $$      {\displaystyle A=1-1+1-1+1-\cdots} $$

We notice that, by reorganisation of the terms of the sum:
\begin{eqnarray} A&=&1-1+1-1+1-1+...\\
&=& 1-(1-1+1-1+...)=1-A \end{eqnarray}

Therefore $$  A={\red {1 \over 2} }\, !!$$
\medskip

$\bullet$ {\bf Determination of $B$}
\medskip

Let's start with
    $$  {\displaystyle B=1-2+3-4+5-6+7-...} $$

We note that by taking the term-by-term difference, we have:

   $$    {\displaystyle {\begin{aligned}B-A&=&1&-2&+3&-4&+5&-6&...&\\&&-1&+1&-1&+1&-1&+1&...&\\&=&0&-1&+2&-3&+4&-5&...&=-B\end{aligned}}}$$

Thus: $$  {\displaystyle 2B=A   \implies }{\displaystyle {\red B= {\frac {1}{4}}}} \, !!
$$

\medskip

$\bullet$ {\bf Determination of $S$}
\medskip

   $$   {\displaystyle S=1+2+3+4+5+...}  $$

We note that by taking the term-by-term difference, we have:

  $$    {\displaystyle {\begin{aligned}S-B&=&1&+2&+3&+4&+5&+6&...&\\&&-1&+2&-3&+4&-5&+6&...&\\&=&0&+4&+0&+8&+0&+12&...&=4 \, S\end{aligned}}}$$

Thus: $$ {\displaystyle S-4S=B}  \implies   {\displaystyle -3S=B}   \implies 
 {\displaystyle {\red S=-{\frac {1}{12}}}}\ !!$$

\section{Poisson summation formula}
\label{app.Poisson}


Consider the series of the form:

\be S_p(x)= \sum_{n=1}^\infty  n^p \ \Theta(x -n)     \ee
with $x>0$. The Poisson summation formula:


\be \sum_{n=0}^\infty f(n)= {1 \over 2} f(0) + \int_0^\infty f(y) dy + 2\sum_{m=1}^\infty \int_0^\infty f(y) \cos(2  \pi m y) d y \ee
gives, with the notation $\omega=2\pi m$:

\be S_0(x)= x+{1 \over 2} +  \sum_{m>0} {1 \over  \pi m} \sin \rr x =x+{1 \over 2} + \osc(x)   \label{S0P} \ee

\be S_1(x)= {\blue{x^2 \over 2}} + 2 \sum_{m=1}^\infty \left[{\cos \rr x {\red - 1} \over \rr^2 }+ {x \sin \rr x \over \rr} \right] = {x^2 \over 2} -{{\red 1 \over 12}} + \osc(x) \label{S1P} \ee

\be S_2(x)= {\blue{x^3 \over 3}} + 2 \sum_{m=1}^\infty \left[{2  \rr x \cos \rr x + (\rr^2 x^2 -2) \sin \rr x \over \rr^3 }\right]={x^3 \over 3} + \osc(x) \label{S2P} \ee

       \begin{eqnarray} &&S_3(x)= {\blue{x^4 \over 4}} + 2 \times \nonumber \\
			&& \sum_{m=1}^\infty \left[{6+ 3 (\rr^2 x^2-2) \cos \rr x + \rr x(\rr^2 x^2 -6) \sin \rr x \over \rr^4 }\right] \nonumber\\
			&&={x^4 \over 4} +{\red {1 \over 120}} + \osc(x)  \label{S3P} \end{eqnarray}

\begin{eqnarray} &&\int_0^x (x^3-y^3) \cos \rr y \, dy \nonumber \\
&&= {{\red -6} + (6- 3 \rr^2 x^2) \cos \rr x + 6 \rr \sin 2 \rr x\over \rr^4}
 \end{eqnarray}

\noindent
More generally the Poisson summation formula leads to (we use complex notation and $\Re$ is the real part. )
\be S_p(x)={x^{p+1} \over p+1}  +  2 \, \Re \sum_{m=1}^\infty { \left[\Gamma(p+1)-\Gamma(p+1, - 2 i m \pi x)\right] \over (- 2 i m \pi)^{p+1}} 
\ee
where $\Re$ is the real part.
The last term, $x$ dependent, contains the oscillations, that we do not consider here. It remains, after extraction of the real part:

\begin{eqnarray}  
\! \!S_p(x)  \! \! \! &=&\!  \! \! {x^{p+1} \over p+1} + {2 \Gamma(p+1) \over (2 \pi)^{p+1}    }   \cos {(p+1) \pi\over 2} \sum_1^\infty {1 \over m^{p+1}} \nonumber \\ \! \! \!&=& \! \! \!
{x^{p+1} \over p+1} + {2  \Gamma(p+1)\over (2 \pi)^{p+1} }  \cos {(p+1) \pi\over 2}\zeta(p+1) \nonumber
\end{eqnarray}
Using the Riemann functional relation \cite{fonctionnelle}, we finally get:
 
\be S_p(x)={\blue {x^{p+1} \over p+1}} + {\red \zeta(-p) }
\ee
as found in (\ref{Spx}). The sum of even powers being zero corresponds to the trivial zeros of the zeta function.

\section{Damping of the oscillations}
\label{app.damping}

Consider the sum

\be S(\ep_F)= \sum_{n= 0}^\infty (n+\varphi)^p \, \Theta(\ep_F - n \omega_c) \label{Sdee0}\ee
It has two energy scales, $\ep_F$ which drives the infinity, and $\omega_c$ which extracts the Ramanujan sum.  We now consider a smooth cut-off. We replace the Heaviside function by a smooth function $\chi_\beta (\ep_F - n \omega_c)$, which varies rapidly on a scale $1/\beta$ much smaller than $\ep_F$. Under this hypothesis, we can replace the new sum 
\be S_\beta(\ep_F)= \sum_{n=0}^\infty n^p \, \chi_\beta(\ep_F - n \omega_c) \ee
by
\be S_\beta(\ep_F)= \sum_{n = 0}^\infty n^p \int \Theta(\ep - n \omega_c) \chi'_\beta(\ep_F - \ep) d \ep \ee
where $\chi'$ is the derivative with respect to $\ep$.
This is done with an integration by parts. The interest of this transformation is to write the new sum in the form
\be S_\beta(\ep_F)=  \int S(\ep) \chi'_\beta(\ep_F - \ep) d \ep \ee
where $S(\ep)$ is given by (\ref{Sdee0}).
After the Poisson transformation of $S(\ep)$, we get terms of the form
\be \int \cos 2 \pi m {\ep \over \omega_c} \, \chi'_\beta(\ep_F - \ep) d \ep  \ee
which may be written in the form
\be R(Z) \ \cos 2 \pi m {\ep_F \over \omega_c} \ee
where $R(Z)$ is a reduction factor ($Z$ is to be explicited below).
\be R = \int \cos 2 \pi m {\delta \ep \over  \omega_c} \chi_\beta'(\delta \ep) d \ep \ee
We have introduced the difference $\delta\ep=\ep_F - \ep$,
\bigskip

Consider first the Fermi-Dirac function as first example of a smooth fonction

\be \chi_\beta(\delta \ep)= {1 \over e^{\beta \delta \ep}+1}  \label{chibeta}\ee
and its derivative is a peaked function around the Fermi energy:
\be \chi'_\beta(\delta \ep)={\beta \over 4 \cosh^2\beta \delta\ep/2 }\ee
The reduction factor is therefore given by

\be R(Z)=\int {\cos z Z  \over 4 \cosh^2 z/2} d z \ee
that is 
\be R(Z)= {\pi Z \over \sinh(\pi Z)}\ee
where $Z= {2 \pi m / \beta \omega_c}$
\medskip

Consider impurity scattering. The smoothed Fermi-Dirac function is now given by\cite{Shoenberg}

\be \chi_\tau(\ep_F - \ep_n)={1 \over 2} + {1 \over \pi} \arctan {(\ep_F - \ep_n)
\over 1/ (2 \tau) } \label{chitau}\ee
where $\tau$ is the scattering time. Its derivative is peaked around the Fermi energy:

\be \chi_\tau'(\delta\ep)= {1/(2 \pi \tau) \over \delta\ep^2 + (1/2 \tau)^2} \ee
and the reduction factor is

\be R(Z)= {1 \over 2 \pi} \int {\cos Z z  \over z^2+1/4}  dz \ee
so that 
\be R(Z)= e^{- Z/2}\ee
with
$Z= 2 m \pi /\omega_c \tau$.

We see that the effect of a smooth cut-off with $1/\beta \ll x$ is to kill the oscillations but does not modify the infinite and the Ramanujan sum. 

In dimensionless units the fonction $\Theta(x-n)$ has to be replaced by a smooth function with varies on a scale larger much smaller than $x$.

 \section{Some precaution about the cut-off}
\label{app.cut-off}

\begin{figure}[h!]
\includegraphics[width=7cm]{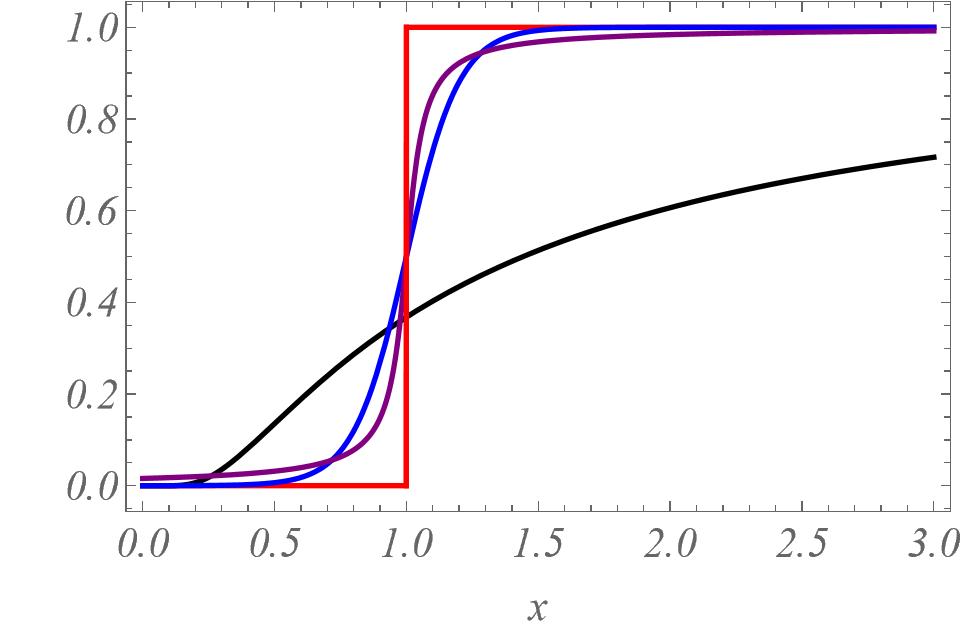}          \\
\includegraphics[width=7cm]{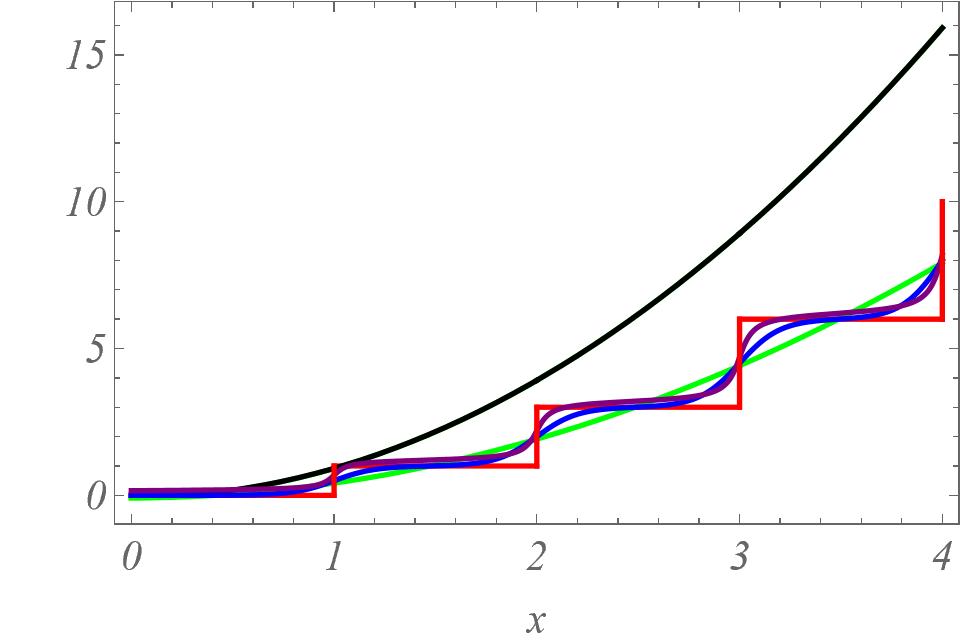}
 \caption{ \small \it Top: different cut-off functions. $\Theta(x-n)$ (red) , $\chi_\beta(x-n)$ (blue), $\chi_\tau(x-n)$ (purple), $e^{-n/x}$ (black). Here we have taken $n=1$, $\beta=\tau=10$. Bottom: sum of integers with the corresponding cut-off function. Obviously the exponential cut-off is not appropriate.}
 \label{fig:cut-off}
\end{figure}

A popular and simple smooth function (regularisation) is sometimes proposed. Instead of the smoothed step-like fonction that we have introduced with an appropriate $\beta$ to suppress the oscillations, an alternative and simple form is simply en exponential $e^{- n /x}$ (Fig. \ref{fig:cut-off}).  It is convenient since the infinite sum can be easily calculated

\be \sum_{n=1}^\infty e^{-n/x}={1 \over 4 \sinh^2 (1/2x)} \ . \ee
In the limit $x \rightarrow \infty$, the sum varies as
\be {x^2  } -{1\over 12}\ee
which is a cheap and easy way to recover the $-1/12$.
\medskip

However, if one is interested on how the sum goes to infinity, the dominant term, the correct result,   must go to infinity as $x^2/2$ and {\it not} $x^2$. Here the first term is meaningless
since it does not tell us correctly how the sum goes to infinity. This is clearly shown in (Fig. \ref{fig:cut-off}).
This is even worse when the power of the naturals increases:

\noindent
Compare
\be \sum_{n=0}^\infty n^3 \chi_\beta(x - n) \longrightarrow {{\blue x^4 \over 4}} +{{\red 1 \over 120}} \nonumber \ee
 with
\be \sum_{n=0}^\infty n^3 e^{-n/x}  \longrightarrow 6 \ x^4  +{{\red 1 \over 120}} \ .\ee
Compare
\be \sum_{n=0}^\infty n^p \chi(x - n) \longrightarrow {{\blue x^{p+1} \over p+1}}+ {\red \zeta(-p)}  \nonumber \ee
with
\be \sum_{n=0}^\infty n^p e^{-n/x}  \longrightarrow   p!\ x^{p+1}  +{\red \zeta(-p)} \ .  \ee

\bigskip

\bibliographystyle{crunsrt}



\end{document}